\documentclass[12pt,english,aps,manuscript]{article}
\usepackage[T1]{fontenc}
\usepackage[latin1]{inputenc}
\usepackage{geometry}
\usepackage[active]{srcltx}
\usepackage{amsmath}
\usepackage{amssymb}
\usepackage[numbers]{natbib}

\makeatletter
\usepackage{geometry}

\makeatletter
\usepackage{color}

\makeatletter
\newcommand{\bee}{\begin{equation}}
\newcommand{\ee}{\end{equation}}
\newcommand{\beea}{\begin{eqnarray}}
\newcommand{\eea}{\end{eqnarray}}

\makeatother

\makeatother

\makeatother

\usepackage{babel}
\begin{document}
\begin{center}
\textbf{\Large{}Wilsonian Effective Field Theory and String Theory{*}}{\Large\par}
\par\end{center}

\begin{center}
\vspace{0.3cm}
\par\end{center}

\begin{center}
{\large{}S. P. de Alwis$^{\dagger}$ }{\large\par}
\par\end{center}

\begin{center}
Physics Department, University of Colorado, \\
 Boulder, CO 80309 USA 
\par\end{center}

\begin{center}
\vspace{0.3cm}
 
\par\end{center}

\begin{center}
\textbf{Abstract} 
\par\end{center}

\begin{center}
\vspace{0.3cm}
\par\end{center}

\smallskip{}
\vspace{0.3cm}
 We argue that deriving an effective field theory from string theory
requires a Wilsonian perspective with a physical cutoff. Employing
proper time regularization we demonstrate the decoupling of states
and contrast this with what happens in dimensional regularization.
In particular we point out that even if the cosmological constant
(CC) calculated from some classical action at some ultra-violet scale
is negative, this does not necessarily imply that the CC calculated
at cosmological scales is also negative, and discuss the possible
criteria for achieving a positive CC starting with a CC at the string/KK
scale which is negative. Obviously this has implications for swampland
claims.

\vfill{}

{*}Contribution to the Peter Suranyi Festschrift $^{\dagger}$ dealwiss@colorado.edu 

\eject

\section{Introduction: EFT from string theory - physical cutoffs vs dimreg}

String theory is supposed to be an ultraviolet complete theory of
quantum gravity. Currently this assertion can be explicitly demonstrated
only in the context of perturbation theory around flat space but it
is widely expected to be valid more generally. This is because in
going from point particles to strings one explicitly introduces (albeit
in a Lorentz invariant fashion) a short distance cutoff - namely the
string length $l_{s}=\sqrt{\alpha'}$ which is a measure of the size
of fundamental strings. Equivalently there is an ultra-violet (UV)
mass scale governed by the tension of the string $T_{s}=1/2\pi\alpha'$,
which is a measure of the mass of the string excitations $M_{s}^{(n)2}=n/\alpha',\,n\in{\cal Z}^{+}$.

In order define the limits of validity of the EFT let us write the
ten-dimensional background metric of string theory as,

\begin{equation}
ds^{2}=G_{MN}dX^{M}dX^{N}=e^{\phi/2}[e^{-6u(x)}g_{\mu\nu}(x)dx^{\mu}dx^{\nu}+e^{2u(x)}\hat{g}_{mn}(y)dy^{m}dy^{n}].\label{eq:10metric}
\end{equation}
Here $\hat{g}_{mn}(y)$ is a fiducial metric on the internal space
$X$ with coordinates $y$ and fiducial volume $\int_{X}\sqrt{\hat{g}}d^{6}y=(2\pi)^{6}\alpha'^{3}$
and $\phi$ is the dilaton. The volume (modulus) of the internal space
is then\footnote{For simplicity we will ignore the effect of warping as we may do for
an internal space of very large volume.} ${\cal V}=e^{6u}$. We have then the relation between the 4D Planck
scale and string scale,

\begin{equation}
T_{s}\equiv M_{s}^{2}=M_{P}^{2}\frac{e^{\phi/2}}{2{\cal V}}.\label{eq:TsMp}
\end{equation}
 Here $M_{P}^{2}=1/8\pi G$ is the low energy Planck scale. The actual
cutoff for a 4D EFT is however below the string scale and is given
by the Kaluza-Klein scale
\begin{equation}
M_{KK}^{2}=\frac{M_{s}^{2}}{{\cal V}^{1/3}}=\frac{M_{P}^{2}}{2{\cal V}^{4/3}}e^{\phi/2}.\label{eq:KK}
\end{equation}
We will focus on type IIB compactifications which are the best studied
in terms of moduli stabilization, phenomenology and cosmology. The
two main scenarios are those of KKLT \citep{Kachru:2003aw} and LVS
\citep{Balasubramanian:2005zx}. In both the dilaton and complex structure
moduli are determined by a set of internal fluxes while the Kaehler
moduli (including the volume ${\cal V}$) are determined by a combination
of internal fluxes and non-perturbative terms (which depend also on
a choice of open string data in particular a gauge group). 

Now given a set of internal space data, i.e. choice of Calabi-Yau
(CY) manifold, internal fluxes and open string data, one has a 4D
EFT with a cutoff essentially given by \eqref{eq:KK}. However it
is possible to keep the field content and the interactions of the
EFT fixed (say by fixing the CY and the open string data) while varying
the fluxes. In the string theory EFT context one can identify this
as a Wilsonian UV cutoff for the EFT. 

On the other hand suppose we have identified a string model in the
landscape of solutions to the string equations, that describes our
universe. At a minimum it must contain the standard model spectrum
and perhaps the spectrum of a supersymmetric extension of it such
as the MSSM. In that case the input for the EFT coming from string
theory would be a set of couplings (including the cosmological constant)
that would be the initial conditions for the RG evolution of the field
theory. These are defined not at some arbitrary adjustable scale but
at a physical scale - the effective Kaluza-Klein scale. In fact this
was pointed out a while ago in \citep{Cicoli:2007xp} where a comparison
of the string loop calculation with the Coleman-Weinberg one-loop
potential with a physical cutoff was made.

This is in contrast to the situation in a purely field theoretic analysis.
In the latter the parameters including the CC are fixed at some scale
through experiments done at that scale. The RG is then used to compute
the effective couplings for processes at a different scale. For instance
one may use the electromagnetic coupling $\alpha$ at very low energies
(say measured in Thompson scattering) and then evolve it up to the
Z-pole to compute QED corrections to processes at that scale. Similarly
with the strong coupling $\alpha_{s}$ which is usually quoted at
the Z-pole, may then be used after RG evolution to compute QCD corrections
at a different scale. 

However none of this depends on the value of those couplings at the
cutoff scale for the standard model coming from the UV completion
of the standard model whatever it is. Indeed any such UV completion
must give values for those couplings which are in agreement with what
has been computed from running the RG up from low energies to the
cutoff scale. What we will find is that the evolution between low
energy and the KK scale will be very different in regularizations
that employ a physical cutoff and in dimensional regularization (dimreg).
The basic reason is that the latter involves integration over all
scales (up to infinite momenta) and the consequent absence of decoupling,
which has to be put in by hand. 

The most salient manifestation of this is in the evolution of the
CC. Whereas with a physical cutoff one sees that there is no significant
evolution up until the lowest mass scale in dimensional regularization
this is not the case and results in a logarithmic infrared divergence
as the scale becomes arbitrarily small. Given that the initial CC
(at the KK scale) is supposed to be finite there is no way to infer
a finite CC at cosmological scales. To put it another way if one inputs
the measured CC at the latter scale, which is the standard prescription
for renormalization in dimreg, then the CC at the UV scale would diverge
- but that would be in conflict with the assumed existence of a finite
UV theory - such as string theory which has to replace the field theory
once one crosses the cutoff value such as the KK scale. In order to
correct this one has, as we mentioned above, to put in the decoupling
of states by hand. 

Furthermore let us suppose that the various no-go theorems which appear
to rule out dS solutions to string theory are actually valid\footnote{This author does not believe this to be the case since this would
mean that all the explicit constructions of a positive CC at the string
scale are invalid for some reason which the advocates of swampland
conjectures have yet to demonstate!}. This would however only mean that at the cutoff scale for the field
theory (presumably some effective KK scale or perhaps the string scale)
there is no dS solution i.e. the CC is either zero or negative or
indeed the potential is runaway. However the observed CC cannot be
compared directly with this string theory calculation since one needs
to take into account the effect of all the low energy fluctuations
below the KK/string scale. In dim reg (at least to one loop) this
appears to yield only negative contributions and hence would not lift
a negative CC to a positive value at cosmological scales. On the contrary
with a physical cutoff depending on the details of low energy physics
there is a possibility, as we shall see later of lifting the low energy
CC to positive values even if the string theory CC is negative\footnote{An example of this possibility has already been given in \citep{deAlwis:2019aud}. }.

\section{Proper time regularization of the quantum effective action}

The quantum theory corresponding to a given classical action $I[\phi]$
is given by the quantum effective action $\Gamma(\phi_{c})$ defined
(implicitly and formally) by the formula\footnote{We work in the Euclidean formulation of quantum field theory. This
is just a matter of convenience. If one worked in the Lorenzian formulation
we would have to include $i\epsilon$ prescriptions and use contour
rotation to define integrals over time components of momenta. The.
expressions for effective actions and RG equations, are of course
not affected.} 
\begin{equation}
e^{-\Gamma(\phi_{c})}=\int[d\phi]e^{-I[\phi]-J.(\phi-\phi_{c})}|_{J=-\partial\Gamma/\partial\phi_{c}}.\label{eq:Gamma}
\end{equation}
By translating the integration variable $\phi=\phi_{c}+\phi'$ we
have the following expressions,
\begin{eqnarray}
e^{-\Gamma(\phi_{c})} & = & \int[d\phi']e^{-I[\phi_{c}+\phi']-J.\phi'}|_{J=-\partial\Gamma/\partial\phi_{c}}\nonumber \\
 & = & \int[d\phi']e^{-\{I[\phi_{c}]+\frac{1}{2}\phi'.\frac{\delta^{2}I}{\delta\phi_{c}^{2}}.\phi'+I_{{\rm i}}[\phi_{c},\phi']+(J+\frac{\delta I[\phi_{c}]}{\delta\phi_{c}}).\phi'\}}|_{J=-\delta\Gamma/\delta\phi_{c}}\label{eq:Gamma2}\\
 & = & e^{-I[\phi_{c}]}e^{-\frac{1}{2}{\rm Trln}K[\phi_{c}]}e^{-I_{{\rm i}}[\phi_{c},-\frac{\delta}{\delta\bar{J}}]}e^{\frac{1}{2}\bar{J}.K[\phi_{c}]^{-1}.\bar{J}}|_{\bar{J}=\delta I[\phi_{c}]/\partial\phi_{c}-\delta\Gamma/\delta\phi_{c}}.\nonumber 
\end{eqnarray}
In the second line above $I_{{\rm i}}[\phi_{c},\phi']$ contains all
powers of $\phi'$ which are higher than quadratic in the expansion
of $I[\phi_{c}+\phi']$, and the third line is the result of doing
the Gaussian integral over $\phi'$. Also $K[\phi_{c}]$ is the kinetic
operator in the presence of the background field $\phi_{c}$ and we've
used a condensed notation so that for example $J.\phi\equiv\int\sqrt{g}\phi^{i}J_{i}$
etc..

Now typically a QFT cannot be assumed to be ultraviolet complete.
Thus the ``initial'' action $I$ in the above is actually some effective
action obtained by integrating out high energy degrees of freedom
(above some UV scale $\Lambda$) from some fundamental UV complete
theory such as string theory. Thus we should replace $I\rightarrow I_{\Lambda}.$
Also the above is a formal expression that needs to be regularized
if we are to evaluate it in perturbation theory. A convenient way
of doing this for our purposes is to introduce the Schwinger proper
time regularization\footnote{In general $K$ will of course be a matrix over space-time indices
as well as internal indices labelling the different fields as well
as their components.} ,
\begin{equation}
{\bf K}_{\Lambda}^{-1}(\phi_{c};x,y)=<x|\int_{1/\Lambda^{2}}^{\infty}dse^{-\hat{{\bf K}}_{\Lambda}[\phi_{c}]s}|y>,\,{\rm ln}{\bf K}_{\Lambda}[\phi_{c};x,y]=-<x|\int_{1/\Lambda^{2}}^{\infty}\frac{ds}{s}e^{-\hat{{\bf K}}_{\Lambda}[\phi_{c}]s}|y>.\label{eq:regulator-1}
\end{equation}
The quantum effective action should be independent of the regularization
scale $\Lambda$. This implies that the ``initial'' action which
one may think of as coming from the low energy limit of some UV complete
theory such as string theory is dependent on the scale $\Lambda$.
i.e. as we stated before, $I\left[\phi\right]\rightarrow I_{\Lambda}[\phi]$
where in the latter all couplings (including masses and the CC) are
$\Lambda$ dependent. Hence the subscript $\Lambda$ on ${\bf K}$
in the RHS of the above equations. The regularized definition of the
1PI action is then obtained by making these replacements in \eqref{eq:Gamma2},

\begin{equation}
e^{-\Gamma\phi_{c})}=e^{-I_{\Lambda}[\phi_{c}]}e^{-\frac{1}{2}{\rm Trln}{\bf K}_{\Lambda}[\phi_{c}]}e^{-I_{{\rm i},\Lambda}[\phi_{c},-\frac{\delta}{\delta\bar{J}}]}e^{\frac{1}{2}\bar{J}.{\bf K}_{\Lambda}[\phi_{c}]^{-1}.\bar{J}}|_{\bar{J}=\delta I_{\Lambda}[\phi_{c}]/\partial\phi_{c}-\delta\Gamma/\delta\phi_{c}}.\label{eq:GammaReg}
\end{equation}
Note that the second exponential factor is the one-loop determinant
and the last two exponentials give the higher than one-loop contributions.
Requiring the independence of $\Gamma$ from the cutoff scale $\Lambda$
i.e. $\Lambda\frac{\partial\Gamma}{\partial\Lambda}=0$ then gives
the set (in general infinite) of RG equations for the couplings in
$I_{\Lambda}$.

Define the heat kernel\footnote{For our purposes here it is sufficient to work in flat background.
The discussion can be easily exted to general curved backgrounds.
A useful review of heat kernel methods is \citep{Vassilevich:2003xt}.
For an application for comparing RG running in higher dimensional
and 4D supergravity see \citep{Hoover:2005uf}.} 
\begin{equation}
{\bf H}(s|x,x')=<x|e^{-{\bf K}s}|x'>,\,<x|x'>=\frac{1}{\sqrt{g}}\delta^{D}(x-x'),\label{eq:heat}
\end{equation}
the formal solution of heat equation $\partial_{s}H(s|x,x')=-{\bf K}H(s|x,x')$.
Note that in general ${\bf K}$ (and hence ${\bf H}$) is a matrix
in internal (field) space as well as in space time. For a scalar field
theory ${\bf K}=-\square{\bf I}+{\bf V}''(\phi_{c})$ where ${\bf V}^{''}$is
the second derivative matrix of the scalar potential. A regularized
one-loop effective action is then given by keeping just the first
two exponential factors of \eqref{eq:GammaReg}, 
\[
\Gamma^{(0+1)}[\phi_{c}]=I_{\Lambda}\left[\phi_{c}\right]+\frac{1}{2}{\rm Trln}{\bf K}_{\Lambda}[\phi_{c}].
\]
The one-loop contribution may be evaluated as,

\begin{eqnarray}
\Gamma^{(1)} & = & \frac{1}{2}{\rm Tr}\ln{\bf K}=-\frac{1}{2}\int_{1/\Lambda^{2}}^{\infty}{\rm Tr}\frac{e^{-{\bf K}[\phi_{c}]s}}{s}ds\nonumber \\
 & = & -\frac{1}{2}\int_{1/\Lambda^{2}}^{\infty}\frac{ds}{s}\int d^{4}x\sqrt{g}{\rm tr}{\bf H}(s|x,x).\label{eq:1-lpEA}\\
 & = & -\frac{1}{2}\int_{1/\Lambda^{2}}^{\infty}\frac{ds}{s}\int d^{4}x\sqrt{g}{\rm tr}e^{-{\bf V}^{''}(\phi_{c})s}\frac{1}{(4\pi s)^{2}}.
\end{eqnarray}
Here ${\bf V}"$ is the mass matrix and we've used the standard result
for the heat kernel of the canonical kinetic (derivative) term in
flat space and ignored space-time variations of the field (and the
metric $g)$, since in this work we will just focus on the effective
potential. 

The expression \eqref{eq:GammaReg} gives a well-defined expression
for the quantum effective action that is obtained by integrating over
all quantum fluctuations down to arbitrarily small energy scales.
It is well-defined in that there is no issue with UV divergences since
it is expressed in terms of the Wilsonian action at the scale $\Lambda$
and the effect of integrating over all quantum fluctuations from $\Lambda$
down to zero. It is assumed that there are no IR divergences or that
they can be effectively taken care of even when there are massless
particles. 

On the other hand one may also define the Wilsonian action at some
lower (non-zero) scale $\mu$ in terms of that at the scale $\Lambda$
by integrating over the quantum fluctuations between those two scales.
For this purpose we may define the propagator and the log of the kinetic
matrix by introducing an IR cutoff into the definitions \eqref{eq:regulator-1},
i.e. 
\begin{equation}
K_{\mu,\Lambda}^{-1}(\phi_{c};x,y)=<x|\int_{1/\Lambda^{2}}^{1/\mu^{2}}dse^{-\hat{K}_{\Lambda}[\phi_{c}]s}|y>,\,{\rm ln}K_{\mu,\Lambda}[\phi_{c};x,y]=-<x|\int_{1/\Lambda^{2}}^{1/\mu^{2}}\frac{ds}{s}e^{-\hat{K}_{\Lambda}[\phi_{c}]s}|y>.\label{eq:regulator2}
\end{equation}
Using these in \eqref{eq:GammaReg} then gives the Wilsonian effective
action at scale $\mu$ in terms of that at the scale $\Lambda$,
\begin{equation}
e^{-I_{\mu}(\phi_{c})}=e^{-I_{\Lambda}[\phi_{c}]}e^{-\frac{1}{2}{\rm Trln}K_{\mu,\Lambda}[\phi_{c}]}e^{-I_{{\rm i},\Lambda}[\phi_{c},-\frac{\delta}{\delta\bar{J}}]}e^{\frac{1}{2}\bar{J}.K_{\mu,\Lambda}[\phi_{c}]^{-1}.\bar{J}}|_{\bar{J}=\delta I_{\Lambda}/\partial\phi_{c}-\delta I_{u}/\delta\phi_{c}}\label{eq:Wilsonmu-Lambda}
\end{equation}
To one-loop we have (upto derivative terms which we ignore since we
are only considering the effective potential below),
\begin{equation}
I_{\mu}(\phi_{c})=I_{\Lambda}[\phi_{c}]+\frac{1}{2}{\rm Trln}K_{\mu,\Lambda}[\phi_{c}]+\ldots=I_{\Lambda}[\phi_{c}]-\frac{1}{2}\int_{1/\Lambda^{2}}^{1/\mu^{2}}\frac{ds}{s}\int d^{4}x\sqrt{g}{\rm tr}e^{-{\bf V}_{\Lambda}^{''}(\phi_{c})s}\frac{1}{(4\pi s)^{2}}+\ldots\label{eq:ImuILambda}
\end{equation}
From this we have the one loop beta function equation\footnote{Using this procedure an exact RG equation was derived in \citep{deAlwis:2017ysy}
and is in effect the RG improved version of the equation below, i.e.
with ${\bf V}_{\Lambda}^{''}(\phi_{c})\rightarrow{\bf V}_{\mu}^{''}(\phi_{c})$.} 
\begin{equation}
\mu\frac{d}{d\mu}I_{\mu}[\phi_{c}]=\frac{\mu^{4}}{16\pi^{2}}\int d^{4}x\sqrt{g}{\rm tr}e^{-{\bf V}_{\Lambda}^{''}(\phi_{c})/\mu^{2}}.\label{eq:floweqn}
\end{equation}

\section{A toy model of decoupling - Heat kernel vs dim reg}

Let us now consider a toy model with low energy degrees of freedom
coupled to high energy modes. Let us see how this happens explicitly
in the context of heat kernel regularization and then compare it to
what one would have done in dimensional regularization.

Let us calculate the $\beta$-function for a scalar field theory with
two scales. The potential is 
\begin{equation}
V(\phi,\Phi)=\Lambda_{{\rm cc}}+\frac{1}{2}m^{2}\phi^{2}+\frac{1}{2}M^{2}\Phi^{2}+\frac{\lambda}{4!}\phi^{4}+\frac{\eta}{4}\phi^{2}\Phi^{2}+\ldots,\label{eq:scalarpot}
\end{equation}
where $M^{2}\gg m^{2}$ and the ellipses stand for higher dimension
operators. The coefficients of the operators in the above are taken
to be defined at some UV scale $\Lambda$ which we may be taken to
be the KK scale in the string theory context. For the purposes of
this discussion we may think of $\phi$ to be the standard model Higgs
with physical (pole) mass $m_{{\rm phy}}=m+O(\lambda,\eta,\ldots)$
and $\Phi$ a Higgs of some grand unified theory with physical (pole)
mass $M_{{\rm phy}}=M+O(\lambda,\eta,\ldots)$. The kinetic operator
in the background $(\phi_{c},0)$ (we set $\Phi$ to zero in order
to focus on the light field beta function) is then $K_{ij}=\square\delta_{ij}+V_{ij}(\phi_{c},0),\,i,j=\phi,\Phi$.
For the above potential we have $V_{\phi\phi}(\phi_{c},0)=m^{2}+\frac{\lambda}{2}\phi_{c}^{2}.\,V_{\Phi,\Phi}=M^{2}+\eta\phi_{c}^{2}$.
Hence from \eqref{eq:floweqn} we have
\begin{equation}
\mu\frac{d}{d\mu}\Gamma^{(1)}=\int d^{4}x\sqrt{g}\frac{\mu^{4}}{16\pi^{2}}[e^{-V_{\phi\phi}/\mu^{2}}+e^{-V_{\Phi\Phi}/\mu{}^{2}}].\label{eq:GdL}
\end{equation}
The effective potential at scale $\mu$ is then given in terms of
the action at scale $\Lambda$ by the relation,
\begin{align}
V^{(0+1)}(\phi,\Phi;\mu) & =\Lambda_{{\rm cc}}(\mu)+\frac{1}{2}m^{2}(\mu)\phi^{2}+\frac{1}{2}M^{2}(\mu)\Phi^{2}+\frac{\lambda(\mu)}{4!}\phi^{4}+\frac{\eta(\mu)}{4}\phi^{2}\Phi^{2}+\ldots\nonumber \\
 & =\Lambda_{{\rm cc}}(\Lambda)+\frac{1}{2}m^{2}(\Lambda)\phi^{2}+\frac{1}{2}M^{2}(\Lambda)\Phi^{2}+\frac{\lambda(\Lambda)}{4!}\phi^{4}+\frac{\eta(\Lambda)}{4}\phi^{2}\Phi^{2}+\nonumber \\
 & +\Delta V^{(1)},\,\,\Delta V^{(1)}=-\frac{1}{2}\int_{1/\Lambda^{2}}^{1/\mu^{2}}\frac{ds}{s}{\rm tr}e^{-{\bf V}^{''}(\phi,\Phi)s}\frac{1}{(4\pi s)^{2}}.\label{eq:VmuVLambda}
\end{align}
Note that in the above we've defined the generalized mass matrix ${\bf V^{''}\equiv}\partial^{2}V/\partial\phi_{i}\partial\phi_{j},\phi_{i}=\phi,\Phi$.
It is important to stress that there is no issue of subtracting divergent
quantities here as is the case in the usual text book discussion of
renormalization in QFT. Equation \eqref{eq:VmuVLambda} relates the
(finite) Wilsonian potential at the scale $\Lambda$ to the (finite)
Wilsonian potential at the scale $\mu$. The one loop contribution
to the relation $V^{(1)}$ is clearly well defined and incorporates
the result of integrating out the degrees of freedom between the two
scales. It can be represented as a difference of two incomplete Gamma
functions.

\begin{align}
\Delta V^{(1)} & =-\frac{1}{2}{\rm tr}\left(V^{''}(\phi_{c})\right)^{2}\int_{V^{''}(\phi_{c)}/\Lambda^{2}}^{V^{''}(\phi_{c})/\mu^{2}}\frac{dt}{t}e^{-t}\frac{1}{(4\pi t)^{2}}\nonumber \\
 & =-\frac{1}{32\pi^{2}}{\rm tr}\left(V^{''}(\phi_{c})\right)^{2}\left[\Gamma\left(-2,{\bf V}^{''}(\phi_{c)}/\Lambda^{2}\right)-\Gamma\left(-2,{\bf V}^{''}(\phi_{c)}/\mu^{2}\right)\right].\label{eq:V1Gamma}
\end{align}
 $\Gamma(\alpha,x)\equiv\int_{x}^{\infty}e^{-t}t^{\alpha-1}$, is
the incomplete Gamma function \footnote{See Gradshteyn and Ryhzik , Tables of Integrals Series and Products
section 8.35 or Abramowitz and Stegan, Handbook of Mathematical functions,
section 6.5. }. For $\alpha=-r,\,r\in{\cal Z}^{+}$ this has the expansion
\begin{align}
\Gamma(-r,x) & =\frac{\left(-1\right)^{r}}{r!}\left[E_{1}(x)-e^{-x}\sum_{j=0}^{r-1}\frac{\left(-1\right)^{j}j!}{x^{j+1}}\right]\label{eq:Gamma-rx}\\
\Gamma(0,x)=E_{1}(x) & \equiv\int_{x}^{\infty}e^{-t}t^{-1}dt=-\gamma-{\rm ln}x-\sum_{n=1}^{\infty}\frac{(-1)^{n}x^{n}}{nn!},\,\,|{\rm arg\,x|<\pi.}\label{eq:E1x}\\
\Gamma\left(-2,x\right) & =\frac{1}{2}\left[E_{1}(x)+e^{-x}\left(\frac{1}{x^{2}}-\frac{1}{x}\right)\right].\nonumber 
\end{align}
In the above $\gamma=.57721...$, is the Euler-Mascheroni constant.
The quantum effective potential up to one loop is then given in this
procedure of regularization by 
\begin{align}
V_{{\rm 1PI}}^{(0+1)}(\phi,\Phi) & =\Lambda_{{\rm cc}}(\Lambda)+\frac{1}{2}m^{2}(\Lambda)\phi^{2}+\frac{1}{2}M^{2}(\Lambda)\Phi^{2}+\frac{\lambda(\Lambda)}{4!}\phi^{4}+\frac{\eta(\Lambda)}{4}\phi^{2}\Phi^{2}+\ldots+V^{(1)}(\phi,\Phi;\Lambda)\label{eq:V1PI}\\
V^{(1)}(\phi,\Phi;\Lambda) & =\frac{1}{32\pi^{2}}{\rm tr}\left({\bf V}^{''}(\phi,\Phi)\right)^{2}\Gamma\left(-2,{\bf V}^{''}(\phi,\Phi)/\Lambda^{2}\right)\nonumber \\
 & =-\frac{1}{64\pi^{2}}{\rm tr}\left[e^{-{\bf V}^{''}/\Lambda^{2}}\left(\Lambda^{4}-\Lambda^{2}{\bf V}^{''}\right)-\left({\bf V}^{''}\right)^{2}\left(\gamma+{\rm ln\frac{{\bf V}^{''}}{\Lambda^{2}}}+\sum_{n=1}^{\infty}\frac{(-1)^{n}}{nn!}\left(\frac{{\bf V}^{''}}{\Lambda^{2}}\right)^{n}\right)\right]\label{eq:V1Lambda}\\
\nonumber 
\end{align}
The RG equations for the local potential are equivalent to the statement
$\Lambda dV_{1{\rm PI}}/d\Lambda=0$. Note again that there is no
issue of infinities here. Given the ``initial'' local potential
(or more generally the action) at the scale $\Lambda$ equation \eqref{eq:V1PI}
gives the effective potential to one loop. Note that up to this point
the discussion is quite general and holds for any scalar field theory.
In the particular case of our toy model the matrix ${\bf V}''$ is
given by \eqref{eq:mass-matrix}.

Although we've focussed on a simple scalar field theory to illustrate
the effects of decoupling (discussed below), we note that the generalization
to include gauge fields and fermions is straightforward - one simply
replaces the traces by supertraces.

How does the above calculation compare to the one in dimreg (reviewed
in the Appendix). The dimreg calculation effectively implies that
all scales are integrated over - so we need to send the cutoff $\Lambda$
to infinity. So in \eqref{eq:V1Lambda} the exponential factor will
tend to unity and the inverse powers of $\Lambda$ will tend to zero.
As for the divergent terms, the potential at $\Lambda$ (usually called
the unrenormalized potential (or action if derivative terms are included)
which is really an completely undefined object with infinite couplings!)
has to be replaced by the so-called renormalized action (with couplings
defined at some arbitrary renormalization scale $\mu$) plus so-called
counter terms (which are of course divergent) chosen to cancel the
divergent one loop terms in the above. Thus for instance the logarithmic
term is rewritten as
\[
\left({\bf V}''\right)^{2}\ln\left(\frac{{\bf V}''}{\Lambda^{2}}\right)=\left({\bf V}''\right)^{2}\ln\left(\frac{{\bf V}''}{\mu^{2}}\right)+\left({\bf V}''\right)^{2}\ln\left(\frac{\mu^{2}}{\Lambda^{2}}\right),
\]
with the second (local) term on the RHS cancelling the counter term
in the ``classical'' action leaving us with just the finite first
term. Thus in this procedure the effective potential to one-loop is
\begin{align*}
V_{{\rm 1PI}}^{(0+1)}(\phi,\Phi) & =\Lambda_{{\rm cc}}(\mu)+\frac{1}{2}m^{2}(\mu)\phi^{2}+\frac{1}{2}M^{2}(\mu)\Phi^{2}+\frac{\lambda(\mu)}{4!}\phi^{4}+\frac{\eta(\mu)}{4}\phi^{2}\Phi^{2}+\ldots+V^{(1)}(\phi,\Phi;\mu)\\
V^{(1)}(\phi,\Phi;\mu) & =\frac{1}{64\pi^{2}}{\rm tr}\left[\left({\bf V}^{''}\right)^{2}\left({\rm ln\frac{{\bf V}^{''}}{\mu^{2}}}+{\rm scheme\,dependent\,finite\,terms}\right)\right]
\end{align*}
in agreement (up to scheme dependence) with the dimreg calculation
of the appendix i.e. eqn. \eqref{eq:1PIdimreg}.

However this interpretation defeats the purpose of having a physical
cutoff. The maximum value of the latter is supposed to be finite since
beyond that value the theory should be replaced by a more fundamental
one which is UV complete as in the case of string theory where necessarily
$\Lambda<M_{{\rm KK}}$. Sending $\Lambda$ to infinity thus makes
no sense and indeed that is why conceptually dimreg is not appropriate
for interpreting the EFT of string theory.

\subsection{The cosmological constant}

First consider the evolution of the cosmological constant $\Lambda_{{\rm cc}}$.
Comparing the coefficients of the unit operator (or $\sqrt{g}$) in
eqn. we have
\begin{align}
\beta_{\Lambda_{{\rm cc}}} & \equiv\mu\frac{d}{d\mu}\Lambda_{{\rm cc}}=\frac{\mu^{4}}{16\pi^{2}}\left[e^{-\frac{m^{2}}{\mu^{2}}}+e^{-\frac{M^{2}}{\mu^{2}}}\right]\label{eq:betaCC0}\\
 & \simeq\frac{\mu^{4}}{16\pi^{2}}\times2,\,\,\,\,\,\,\mu\gg M\gg m\label{eq:betaCC1}\\
 & \simeq\frac{\mu^{4}}{16\pi^{2}},\,\,\,\,\,\,\,\,\,\:m\ll\mu\ll M\label{eq:betaCC2}\\
 & \simeq0,\,\,\,\,\,\,\,\,\,\,\,\,\,\,\,\,\mu\ll m\ll M\label{eq:betaCC3}
\end{align}
Here for the purposes of illustration we have made a crude approximation
in the last three equations which will be refined in the next subsection. 

What do these equations imply for the low energy parameters that should
be used to discuss physics at say the standard model scale or below,
assuming that the initial values are set by some UV complete theory
such as string theory. Given that the latter is the only theory of
quantum gravity that can even in principle include the standard model
we will focus on it.

In string theory there are two scales - one the string scale $M_{s}$
is the mass of the lowest string excitations (and defines the tension
of the string). This controls the low energy expansion of the string
field theory. This is a ten dimensional local field theory with an
infinite number of terms with the coefficients of higher dimensional
operators controlled by inverse powers of $M_{{\rm s}}^{2}$. The
four dimensional theory is obtained by compactifying six of the spatial
dimensions, and that introduces another scale - the Kaluza-Klein scale
$M_{{\rm KK}}$, whose inverse sets the length scale of the compactified
space. It is the latter which controls the low energy expansion of
the four dimensional theory since in proceeding via the ten-dimensional
field theory we've already assumed that the volume of the extra six
dimensions is large and hence for consistency we need to have $M_{{\rm KK}}<M_{{\rm s}}$.
We will assume that the theory we are discussing comes from a particular
point in the landscape of string theory. This means that the initial
data for the RG evolution of the parameters of the field theory are
fixed by string theory.

For our toy model this means that the initial potential) takes the
form \eqref{eq:scalarpot} with parameters
\[
\Lambda_{{\rm cc}}\rightarrow\Lambda_{0}\equiv\Lambda_{{\rm cc}}(M_{KK}),\,m\rightarrow m_{0}\equiv m(M_{KK}),\,M\rightarrow M_{0}\equiv M(M_{KK}),\ldots.
\]
These are all determined by string theory and the corresponding action
may be used to compute physical processes at energies just below the
KK scale. The RG equations then determine the effective low energy
action that may be used at energy scales well below the KK scale. 

We integrate in stages first following the simplest approximations
to \eqref{eq:betaCC0}, i.e. \eqref{eq:betaCC1},\eqref{eq:betaCC2}
and \eqref{eq:betaCC3}. From the first we get 
\[
\Lambda_{{\rm cc}}(M)=\Lambda_{0}+\frac{1}{16\pi^{2}}\frac{1}{2}\left(M^{4}-M_{{\rm KK}}^{4}\right).
\]
From the second we have
\[
\Lambda_{{\rm cc}}(m)=\Lambda_{{\rm cc}}(M)+\frac{1}{16\pi^{2}}\frac{1}{4}\left(m^{4}-M^{4}\right).
\]
 Finally from the third we have 
\[
\Lambda_{{\rm cc}}(\mu\ll m)=\Lambda_{{\rm cc}}(m)
\]
reflecting the fact that the CC hardly evolves between cosmological
scales and the lightest (non-zero) physical mass scale. Putting these
equations together we can express the cosmological CC in terms of
the CC coming from string theory,
\begin{equation}
\Lambda_{{\rm cc}}(\mu\ll m)=\Lambda_{{\rm 0}}+\frac{1}{64\pi^{2}}\left(m^{4}-M^{4}\right)+\frac{1}{32\pi^{2}}\left(M^{4}-M_{{\rm KK}}^{4}\right).\label{eq:CCcosmo}
\end{equation}
This is a fairly crude approximation since we ignored the regions
where the exponential terms are different from zero or one. Nevertheless
this equation nicely illustrates the fine tuning problem for the CC,
at least in a non-supersymmetric theory or a string theory with a
SUSY breaking scale above the KK scale. The input CC coming from string
theory at the KK scale, namely $\Lambda_{0}$, has to be such as to
account for each mass threshold that is crossed as one goes from this
scale down to cosmological scales. The argument of Bousso and Polchinski
\citep{Bousso:2000xa} is that given a sufficiently complicated compactification
manifold (a Calabi-Yau space with a large number of cycles for instance
with internal fluxes turned on) this is always possible.

Let us now include correction terms coming from the exponentials in
\eqref{eq:betaCC0}. In the region $m^{2}\ll M^{2}<\mu^{2}<M_{{\rm KK}}^{2}$,
\begin{align*}
\mu\frac{d}{d\mu}\Lambda_{{\rm cc}} & \simeq\frac{\mu^{4}}{16\pi^{2}}\left[1+\left(1-\frac{M^{2}}{\mu^{2}}+\frac{1}{2}\frac{M^{4}}{\mu^{4}}+O\left(\frac{M^{6}}{\mu^{6}}\right)\right)\right]\\
 & =\frac{1}{16\pi^{2}}\left[2\mu^{4}-M^{2}\mu^{2}+\frac{1}{2}M^{4}+\ldots\right].
\end{align*}
Integrating this we get
\begin{equation}
\Lambda_{{\rm cc}}(M)-\Lambda_{{\rm cc}}(M_{{\rm KK}})\simeq\frac{M^{4}-M_{{\rm KK}}^{4}}{32\pi^{2}}+\frac{M_{{\rm KK}}^{2}-M^{2}}{32\pi^{2}}M^{2}+\frac{1}{64\pi^{2}}M^{4}\ln\left(\frac{M^{2}}{M_{{\rm KK}}^{2}}\right)+\ldots\label{eq:CCMMkk}
\end{equation}
For $m^{2}<\mu^{2}\ll M^{2}$ we have 
\begin{align*}
\mu\frac{d}{d\mu}\Lambda_{{\rm cc}} & \simeq\frac{\mu^{4}}{16\pi^{2}}\left(e^{-m^{2}/\mu^{2}}+0\right)\\
 & =\frac{\mu^{4}}{16\pi^{2}}\left(1-\frac{m^{2}}{\mu^{2}}+\frac{1}{2}\left(\frac{m^{2}}{\mu^{2}}\right)^{2}+\ldots\right)
\end{align*}
Integrating this we get
\begin{equation}
\Lambda_{{\rm cc}}(m)-\Lambda_{{\rm cc}}(M)\simeq\frac{m^{4}-M^{4}}{64\pi^{2}}+\frac{\left(M^{2}-m^{2}\right)}{32\pi^{2}}m^{2}+\frac{m^{4}}{64\pi^{2}}\ln\left(\frac{m^{2}}{M^{2}}\right)+\ldots\label{eq:CCmM}
\end{equation}
Finally in the region $\mu^{2}\ll m^{2}$, we have 
\begin{equation}
\mu\frac{d}{d\mu}\Lambda_{{\rm cc}}\simeq0,\Rightarrow\Lambda_{{\rm cc}}(\mu\ll m)\simeq\Lambda_{{\rm cc}}(m)\label{eq:CC0m}
\end{equation}
Collecting the three expressions we have for the long distance CC
the expression
\begin{align}
\Lambda_{{\rm CC}}(\mu & \ll m)=\Lambda_{{\rm CC}}(M_{{\rm KK}})-\frac{M_{{\rm KK}}^{4}-m^{4}}{32\pi^{2}}+\frac{M_{{\rm KK}}^{2}-M^{2}}{32\pi^{2}}M^{2}+\frac{\left(M^{2}-m^{2}\right)}{32\pi^{2}}m^{2}\nonumber \\
 & +\frac{M^{4}}{64\pi^{2}}\ln\left(\frac{M^{2}}{M_{{\rm KK}}^{2}}\right)+\frac{m^{4}}{64\pi^{2}}\ln\left(\frac{m^{2}}{M^{2}}\right)+\ldots\label{eq:CC0KK}
\end{align}

Before we go on to discuss the evolution of the other parameters,
let us generalize these formulae to a SUSY theory where the supersymmery
is broken below the KK scale. In this case in the formula for the
effective action \eqref{eq:1-lpEA} the trace instruction will be
replaced by the supertrace defined (with ${\cal M}$ being the field
dependent mass matrix) by 
\[
{\rm Str}\left({\cal M}^{2}\right)^{n}=\sum_{j}(2j+1)(-1)^{j}{\rm tr}\left({\cal M}_{j}^{2}\right)^{n}.
\]
Let us take again a hierarchy of mass scales (corresponding for instance
to a superGUT model) with a high scale supermultiplet at a mass scale
$\bar{M}$ and a low scale one at a sale $\bar{m}$. Let us also denote
the corresponding supermultiplet mass matrices as ${\bf M}$ and ${\cal {\bf m}}$.
Equation \eqref{eq:CCMMkk} is then replaced by 
\begin{equation}
\Lambda_{{\rm cc}}(\bar{M})-\Lambda_{{\rm cc}}(M_{{\rm KK}})\simeq\frac{\bar{M}^{4}-M_{{\rm KK}}^{4}}{32\pi^{2}}{\rm Str}{\bf {\bf \left(M^{2}\right)}}^{0}-\frac{\bar{M}^{2}-M_{{\rm KK}}^{2}}{32\pi^{2}}{\rm Str}{\bf M}^{2}+\frac{1}{64\pi^{2}}{\rm Str}{\bf M}^{4}\ln\left(\frac{M^{2}}{M_{{\rm KK}}^{2}}\right)+\ldots,\label{eq:CCMKKsusy}
\end{equation}
and eqn. \eqref{eq:CCmM} by, 
\begin{equation}
\Lambda_{{\rm cc}}(m)-\Lambda_{{\rm cc}}(M)\simeq\frac{\bar{m}^{4}-\bar{M}^{4}}{64\pi^{2}}{\rm Str}{\bf {\bf \left(m^{2}\right)}}^{0}-\frac{\bar{m}^{2}-\bar{M}^{2}}{32\pi^{2}}{\rm Str}{\bf m^{2}}+\frac{1}{64\pi^{2}}{\rm Str}{\bf m}^{4}\ln\left(\frac{\bar{m}^{2}}{\bar{M}^{2}}\right)+\ldots.\label{eq:CCmMsusy}
\end{equation}
For a theory in which there is an equal number of fermionic and bosonic
degrees of freedom and hence even for a broken supersymmetric theory,
${\rm Str}{\bf {\bf \left(M^{2}\right)}}^{0}={\rm Str}{\bf {\bf \left(m^{2}\right)}}^{0}=0$.
Also of course \eqref{eq:CC0m} is unchanged. Hence we have for the
low energy cosmological constant in a SUSY theory,

\begin{align}
\Lambda_{{\rm cc}}(\mu & \ll m)\simeq\Lambda_{{\rm cc}}(M_{{\rm KK}})-\frac{\bar{M}^{2}-M_{{\rm KK}}^{2}}{32\pi^{2}}{\rm Str}_{\bar{M}}{\bf M}^{2}-\frac{\bar{m}^{2}-\bar{M}^{2}}{32\pi^{2}}{\rm Str}_{\bar{m}}{\bf m^{2}}\nonumber \\
 & +\frac{1}{64\pi^{2}}{\rm Str}_{\bar{M}}{\bf M}^{4}\ln\left(\frac{\bar{M}^{2}}{M_{{\rm KK}}^{2}}\right)+\frac{1}{64\pi^{2}}{\rm Str}_{\bar{m}}{\bf m}^{4}\ln\left(\frac{\bar{m}^{2}}{\bar{M}^{2}}\right)+\ldots.\label{eq:CC0susy}
\end{align}
The subscript on the the supertrace instruction implies that it is
to be taken over the supermultiplets at that scale. So for example
in a superGUT theory the subscript $\bar{M}$ implies the supertrace
over the GUT scale supermultiplets and the subscript $\bar{m}$ implies
the supertrace over the MSSM supermultiplets. It is assumed also that
the splitting within a multiplet is much smaller than $\bar{M}-\bar{m}$.

Note that the second and third terms on the RHS of \eqref{eq:CC0susy}
are in fact positive since typically the ${\rm Str}M^{2}(m^{2})$
is positive. On the other hand the third and fourth terms are negative.
The expression shows that whether the quantum correction to the ``classical''
CC generated by string theory (i.e. the initial condition for the
evolution of the CC) does not necessarily have to be positive (see
also \citep{deAlwis:2019aud}). Depending on the physics below the
KK scale, it may be the case that the final CC (at cosmological scales)
can indeed be positive even if the string theory generated CC is negative
at the KK scale. For instance one could take the LVS minimum (before
the so-called uplift which is less well established) with a negative
CC albeit with broken SUSY. In the case of KKLT however this is somewhat
more problematic since the minimum before ``uplift'' is a SUSY preserving
AdS space. 

Let us now compare this formula (i.e. \eqref{eq:CC0susy}) with what
is obtained in dimensional regularization which gives (see Appendix
eqn. \eqref{eq:LambdaCCdimreg})

\begin{equation}
\Lambda_{{\rm cc}}(\mu\ll m)=\Lambda_{{\rm cc}}\left(M_{{\rm KK}}\right)+\frac{1}{64\pi^{2}}{\rm Str}\left({\bf m}^{4}+{\bf M}^{4}\right){\rm ln}\left(\frac{\mu^{2}}{M_{{\rm KK}}^{2}}\right)\label{eq:CCdimreg}
\end{equation}
The decoupling that is manifest in \eqref{eq:CC0susy} is absent in
the above - one needs to put it in by hand. In fact as a consequence
it appears that this expression has an infrared divergence as $\mu\rightarrow0$!
Now if we did not have an UV complete theory one usually thinks of
the cutoff (which is here a physical scale) as a scale which is at
the end of the day sent to infinity. $\Lambda_{{\rm cc}}\left(M_{{\rm KK}}\right)$
would then be thought of as the ``bare'' CC which has no physical
significance. The only number we have is the measured large distance
CC i.e. the LHS of \eqref{eq:CC0susy}\eqref{eq:CCdimreg}. The difference
in the RHS's of two formulae is of no consequence and just means that
the counter terms in the two schemes are different. In the first case
one would need to subtract a quadratic (in a non-SUSY theory also
a quartic) divergence as well as a log divergence whereas in the dim
reg case only a log divergence would be subtracted. However once we
have a meaningful UV complete theory such as string theory the initial
value is not divergent and (for a given point in the landscape) has
a well defined value. This means that only a physical cutoff scheme
makes sense in this context.

\subsection{Masses and couplings}

Let us now discuss the evolution of the coupling $\lambda$. Using
\eqref{eq:GdL} and comparing the coefficients of $\phi_{c}^{4}$
(for instance) on both sides we have ,
\begin{equation}
\beta_{\lambda}\equiv\mu\frac{d}{d\mu}\lambda=\frac{3\lambda^{2}}{16\pi^{2}}e^{-\frac{m^{2}}{\mu^{2}}}+\frac{3\eta^{2}}{16\pi^{2}}e^{-\frac{M^{2}}{\mu^{2}}}.\label{eq:betalam}
\end{equation}
Again we see very clearly from the heat kernel method how states decouple.
Thus we have the following expressions for the beta function in three
different regimes 
\begin{eqnarray}
\beta_{\lambda} & \simeq & \frac{3\lambda^{2}}{16\pi^{2}}+\frac{\eta^{2}}{32\pi^{2}},\,\mu\gg M\gg m.\label{eq:betal1}\\
 & \simeq & \frac{3\lambda^{2}}{16\pi^{2}},\,M\gg\mu\gg m\label{eq:betal2}\\
 & \simeq & 0,\,m\gg\mu.\label{eq:betal3}
\end{eqnarray}
This is in contrast to the result in dimensional regularization where
the decoupling has to be introduced by hand. In fact if one did the
usual calculation in dimreg the answer is just the first line above
\eqref{eq:betal1} as we've reviewed in the Appendix (see eqn.\eqref{eq:lambdaRGdimreg}). 

If one ignores the evolution of $\eta$ these equations can be integrated.
Thus we have, starting with \eqref{eq:betal3}

\begin{align}
\lambda(m) & \simeq\lambda(\mu),\,\mu\ll m\label{eq:la3}\\
\lambda(\mu) & =\frac{\lambda(m)}{1-\frac{3}{16\pi^{2}}\lambda(m)\ln\frac{\mu}{m}},\,,\mu\ll M<m\exp\left[\frac{16\pi^{2}}{3\lambda(m)}\right]\label{eq:la2}\\
\lambda(\mu) & =\frac{\eta}{\sqrt{6}}\tan\left(\frac{3}{16\pi^{2}}\frac{\eta}{\sqrt{6}}{\rm ln}\frac{\mu}{M}+\tan^{-1}\frac{\sqrt{6}\lambda(M)}{\eta}\right),\,M_{{\rm KK}}\gg\mu>M\label{eq:la1}
\end{align}
We note in passing that in order to see how the last equation reduces
to the second in the limit $\eta\rightarrow0$ (as it should) we have
to use the Taylor series expansion around infinity for the inverse
tangent $\tan^{-1}x=\frac{\pi}{2}-\frac{1}{x}+\ldots$. Also in the
second equation the last condition is the requirement that the high
mass threshold must be crossed before the Landau pole. A similar condition
holds for the third equation - namely that $M_{{\rm KK}}$ should
be less than the pole on the RHS of that equation when the argument
of the tangent hits $\pi/2$.

Let us consider now the beta function for the light mass,
\begin{equation}
\beta_{m^{2}}\equiv\mu\frac{d}{d\mu}m^{2}=-\frac{\lambda}{16\pi^{2}}\mu^{2}e^{-\frac{m^{2}}{\mu^{2}}}-\frac{\eta}{16\pi^{2}}\mu^{2}e^{-\frac{M^{2}}{\mu^{2}}}.\label{eq:betam}
\end{equation}
Then in the three regimes we have
\begin{align*}
\beta_{m^{2}} & =-\frac{1}{16\pi^{2}}\left(\lambda+\eta\right)\mu^{2}+\frac{1}{16\pi^{2}}\left(\lambda m^{2}+\eta M^{2}\right)+O(M^{2}/\mu^{2}),\,\mu\gg M\gg m,\\
 & \simeq-\frac{\lambda}{16\pi^{2}}\mu^{2}+\frac{1}{16\pi^{2}}\lambda m^{2},\,m\ll\mu\ll M,\\
 & \simeq0,\,\mu\ll m.
\end{align*}
 On the other hand what happens in dimreg is that one gets the first
line of the above set of eqns without the first term since that corresponds
to a quadratic divergence (in the usual non-Wilsonian discussion)
that is absent in dimreg. i.e. we have 
\begin{equation}
\beta_{m^{2}}=\frac{1}{16\pi^{2}}(\lambda m^{2}+\eta M^{2}),\label{eq:betamdimreg}
\end{equation}
and as with the dimreg eqn. for $\lambda$ this is valid at all scales. 

But again one would not see the decoupling of the heavy states that
is manifest in the second and third lines above. This phenomenon in
dimreg for the coulings the masses and the CC is not surprising, since
dimereg involves integration over all scales and decoupling has to
be introduced by hand\footnote{For a very clear recent discussion of this see \citep{Burgess:2020tbq}
section 7.2.3. }.

\section{Conclusions}

The question we've addressed in this note is the evolution of the
couplings down to long distance scales when the initial (``classical'')
action is given by the effective field theory of some UV complete
theory such as string theory. This EFT is expected to be valid up
to the Kaluza-Klein scale if the UV theory is string theory. We have
argued that in this case it makes more sense to use a physical cut-off
when defining the Wilsonian action for scales well below the KK scale.
The same is the case when obtaining the 1PI action which is supposed
to incorporate all quantum fluctuations i.e. to arbitrarily low scales.
It is defined by an initial ``classical'' action at the KK scale
and then integrating all quantum fluctuations down to the long wave
length limit to get the quantum effective action. 

As we argued earlier, while in a purely field theoretic scenario (say
with just renormalizable couplings) the issue of an initial action
as a meaningful entity does not arise since it is essentially a cut
off dependent object called the bare or unrenormalized action which
in the limit when the cut off is removed goes to infinity. In fact
in dimreg the analog of $\int d^{D}x$ for negative dimension (since
the quartic divergence in the CC can only be regularized for negative
$D$) is meaningless. Actually the situation is even worse since the
procedure needs to be well defined for all real values of $D$ from
negative values to $D=4$. Thus a regularized all orders expression
such as \eqref{eq:GammaReg} which was possible in heat kernel regularization
(or more generally in any such physical regulator scheme) makes no
sense in dimreg.

Nevertheless dimreg remains by far the easiest way of calculating
in perturbation theory beyond one-loop. The issue we have presented
is a conceptual one. Once we have an initial ``classical'' action
coming from a UV complete theory such as string theory a physical
cutoff gives us a clear way of relating low energy physics to the
UV theory. Furthermore this gives an explicit understanding of decoupling
and most importantly shows us that depending on the nature of the
physics in between the UV cutoff (KK) scale and the cosmological scale,
a negative CC at the string/KK scale may still yield the observed
positive CC. This would relieve the current tension between the observed
positive CC at cosmological scales and the CC from string constructions
which is typically negative.

\section{Acknowledgements}

I wish to thank Cliff Burgess, Roberto Percacci and Fernando Quevedo
for comments on the draft. I'm happy to be able to contribute to the
Festschrift celebrating Peter Suranyi's many contributions to a variety
of topics in high energy physics.

\section*{Appendix: Coleman-Weinberg one-loop potential in dimreg}

Calculating with a a momentum space cutoff $\Lambda$ one gets (see
for example \citep{Ferrara:1994kg})

\begin{equation}
V=V_{0}+\frac{1}{64\pi^{2}}{\rm Str}{\cal M}^{0}\Lambda^{4}\ln\frac{\Lambda^{2}}{\mu^{2}}+\frac{1}{32\pi^{2}}{\rm Str}{\cal M}^{2}\Lambda^{2}+\frac{1}{64\pi^{2}}{\rm Str}{\cal M}^{4}\ln\frac{{\cal M}^{2}}{\Lambda^{2}}+\ldots\label{eq:CWpot}
\end{equation}

\[
{\rm Str}{\cal M}^{2n}=\sum_{J}(2J+1)(-1)^{2J}M_{j}^{2n}
\]
 This is the same as our heat kernal regularized expression \eqref{eq:V1PI}\eqref{eq:V1Lambda}
when $\Lambda^{2}\gg{\bf |V}"[\phi]|$. On the other hand in dimreg
only the fourth term is present. For completeness we review the calculation
below. 

In dimensional regularization one may start with the D dimensional
version of the proper time representation eqn. \eqref{eq:1-lpEA}
but without the cut-off in the $s$-integral, i.e.

\[
\Gamma_{D}^{(1)}=-\frac{1}{2}\int d^{D}x\sqrt{g}\int_{0}^{\infty}\frac{ds}{s}{\rm tr}e^{-{\bf V}^{''}(\phi_{c})s}\frac{1}{(4\pi s)^{D/2}}
\]
While the space-time integral makes no sense for negative or non-integral
$D$, the $s$intgral is well-defined for $D<0$. Introducing the
arbitrary mass scale $\mu$ and using the integral representation
for the Gamma function $\Gamma(z)=\int_{0}^{\infty}e^{-t}t^{z-1}dt$
and writing $D=4-\epsilon$ and expanding in a Laurent series in $\epsilon$
we get (see for example Peskin and Schroeder \citep{Peskin:1995ev}
eqns. (11.77,78))
\[
\Gamma_{4-\epsilon}^{(1)}=-\frac{1}{2}\int d^{4}x\sqrt{g}{\rm tr}\frac{\left({\bf V}''\right)^{2}}{(4\pi)^{2}}\frac{1}{2}\left(\frac{2}{\epsilon}-\gamma+\ln(4\pi)-\ln\frac{{\bf V}''}{\mu^{2}}+\frac{3}{2}+O(\epsilon)\right),
\]
 where $\mu$ is an arbitrary scale factor. In $\overline{{\rm MS}}$
one adds the counter term 
\[
\delta S=\frac{1}{2}\int d^{4}x\sqrt{g}{\rm tr}\frac{\left({\bf V}''\right)^{2}}{(4\pi)^{2}}\frac{1}{2}\left(\frac{2}{\epsilon}-\gamma+\ln(4\pi)\right),
\]
to the original action (with couplings defined at the mass scale $\mu$)
so that we have for the one-loop corrected quantum effective action
(1PI action to one-loop)
\begin{align}
\Gamma_{{\rm 1PI}} & \simeq S_{{\rm cl(\mu)}}+\text{\ensuremath{\lim_{\epsilon\rightarrow0}\left(\delta S+\Gamma_{4-\epsilon}^{(1)}\right)}}\nonumber \\
 & =S_{{\rm cl(\mu)}}+\frac{1}{2}\int d^{4}x\sqrt{g}{\rm tr}\frac{\left({\bf V}''\right)^{2}}{(4\pi)^{2}}\frac{1}{2}\left(\ln\frac{{\bf V}''}{\mu^{2}}-\frac{3}{2}\right)\label{eq:1PIdimreg}
\end{align}
In our toy model the classical potential is,
\[
V(\phi,\Phi;\mu)=\Lambda_{{\rm cc}}(\mu)+\frac{1}{2}m^{2}(\mu)\phi^{2}+\frac{1}{2}M^{2}(\mu)\Phi^{2}+\frac{\lambda(\mu)}{4!}\phi^{4}+\frac{\eta(\mu)}{4}\phi^{2}\Phi^{2}+\ldots.
\]
The $\beta$-functions are obtained by demanding that $\Gamma_{{\rm 1PI}}$
is independent of the arbitrary scale $\mu$. Thus we may read off
the flow equations for the couplings:
\begin{align}
\mu\frac{d\Lambda_{{\rm cc}}}{d\mu} & =\frac{1}{32\pi^{2}}\left(m^{4}+M^{4}\right)\label{eq:CCRGdimreg}\\
\mu\frac{dm^{2}}{d\mu} & =\frac{1}{16\pi^{2}}\left(\lambda m^{2}+\eta M^{2}\right)\label{eq:m2RGdimreg}\\
\mu\frac{d\lambda(\mu)}{d\mu} & =\frac{3}{16\pi^{2}}\left(\left(\lambda^{2}+\eta^{2}\right)\right)\label{eq:lambdaRGdimreg}
\end{align}

Let us focus on the cosmological constant. Integrating the first equation
between $M_{{\rm KK}}$ and cosmological scales $\mu\ll m$ we get
(after generalizing to include also fermions and gauge bosons) 
\begin{equation}
\Lambda_{{\rm cc}}(\mu\ll m)=\Lambda_{{\rm cc}}\left(M_{{\rm KK}}\right)+\frac{1}{64\pi^{2}}{\rm Str}\left({\bf m}^{4}+{\bf M}^{4}\right){\rm ln}\left(\frac{\mu^{2}}{M_{{\rm KK}}^{2}}\right).\label{eq:LambdaCCdimreg}
\end{equation}

\bibliographystyle{apsrev}
\bibliography{myrefs}

\begin{thebibliography}{11}
\expandafter\ifx\csname natexlab\endcsname\relax\def\natexlab#1{#1}\fi
\expandafter\ifx\csname bibnamefont\endcsname\relax
  \def\bibnamefont#1{#1}\fi
\expandafter\ifx\csname bibfnamefont\endcsname\relax
  \def\bibfnamefont#1{#1}\fi
\expandafter\ifx\csname citenamefont\endcsname\relax
  \def\citenamefont#1{#1}\fi
\expandafter\ifx\csname url\endcsname\relax
  \def\url#1{\texttt{#1}}\fi
\expandafter\ifx\csname urlprefix\endcsname\relax\def\urlprefix{URL }\fi
\providecommand{\bibinfo}[2]{#2}
\providecommand{\eprint}[2][]{\url{#2}}

\bibitem[{\citenamefont{Kachru et~al.}(2003)\citenamefont{Kachru, Kallosh,
  Linde, and Trivedi}}]{Kachru:2003aw}
\bibinfo{author}{\bibfnamefont{S.}~\bibnamefont{Kachru}},
  \bibinfo{author}{\bibfnamefont{R.}~\bibnamefont{Kallosh}},
  \bibinfo{author}{\bibfnamefont{A.~D.} \bibnamefont{Linde}}, \bibnamefont{and}
  \bibinfo{author}{\bibfnamefont{S.~P.} \bibnamefont{Trivedi}},
  \bibinfo{journal}{Phys. Rev. D} \textbf{\bibinfo{volume}{68}},
  \bibinfo{pages}{046005} (\bibinfo{year}{2003}), \eprint{hep-th/0301240}.

\bibitem[{\citenamefont{Balasubramanian
  et~al.}(2005)\citenamefont{Balasubramanian, Berglund, Conlon, and
  Quevedo}}]{Balasubramanian:2005zx}
\bibinfo{author}{\bibfnamefont{V.}~\bibnamefont{Balasubramanian}},
  \bibinfo{author}{\bibfnamefont{P.}~\bibnamefont{Berglund}},
  \bibinfo{author}{\bibfnamefont{J.~P.} \bibnamefont{Conlon}},
  \bibnamefont{and} \bibinfo{author}{\bibfnamefont{F.}~\bibnamefont{Quevedo}},
  \bibinfo{journal}{JHEP} \textbf{\bibinfo{volume}{03}}, \bibinfo{pages}{007}
  (\bibinfo{year}{2005}), \eprint{hep-th/0502058}.

\bibitem[{\citenamefont{Cicoli et~al.}(2008)\citenamefont{Cicoli, Conlon, and
  Quevedo}}]{Cicoli:2007xp}
\bibinfo{author}{\bibfnamefont{M.}~\bibnamefont{Cicoli}},
  \bibinfo{author}{\bibfnamefont{J.~P.} \bibnamefont{Conlon}},
  \bibnamefont{and} \bibinfo{author}{\bibfnamefont{F.}~\bibnamefont{Quevedo}},
  \bibinfo{journal}{JHEP} \textbf{\bibinfo{volume}{01}}, \bibinfo{pages}{052}
  (\bibinfo{year}{2008}), \eprint{0708.1873}.

\bibitem[{\citenamefont{de~Alwis et~al.}(2019)\citenamefont{de~Alwis, Eichhorn,
  Held, Pawlowski, Schiffer, and Versteegen}}]{deAlwis:2019aud}
\bibinfo{author}{\bibfnamefont{S.}~\bibnamefont{de~Alwis}},
  \bibinfo{author}{\bibfnamefont{A.}~\bibnamefont{Eichhorn}},
  \bibinfo{author}{\bibfnamefont{A.}~\bibnamefont{Held}},
  \bibinfo{author}{\bibfnamefont{J.~M.} \bibnamefont{Pawlowski}},
  \bibinfo{author}{\bibfnamefont{M.}~\bibnamefont{Schiffer}}, \bibnamefont{and}
  \bibinfo{author}{\bibfnamefont{F.}~\bibnamefont{Versteegen}},
  \bibinfo{journal}{Phys. Lett. B} \textbf{\bibinfo{volume}{798}},
  \bibinfo{pages}{134991} (\bibinfo{year}{2019}), \eprint{1907.07894}.

\bibitem[{\citenamefont{Vassilevich}(2003)}]{Vassilevich:2003xt}
\bibinfo{author}{\bibfnamefont{D.~V.} \bibnamefont{Vassilevich}},
  \bibinfo{journal}{Phys. Rept.} \textbf{\bibinfo{volume}{388}},
  \bibinfo{pages}{279} (\bibinfo{year}{2003}), \eprint{hep-th/0306138}.

\bibitem[{\citenamefont{Hoover and Burgess}(2006)}]{Hoover:2005uf}
\bibinfo{author}{\bibfnamefont{D.}~\bibnamefont{Hoover}} \bibnamefont{and}
  \bibinfo{author}{\bibfnamefont{C.~P.} \bibnamefont{Burgess}},
  \bibinfo{journal}{JHEP} \textbf{\bibinfo{volume}{01}}, \bibinfo{pages}{058}
  (\bibinfo{year}{2006}), \eprint{hep-th/0507293}.

\bibitem[{\citenamefont{de~Alwis}(2018)}]{deAlwis:2017ysy}
\bibinfo{author}{\bibfnamefont{S.~P.} \bibnamefont{de~Alwis}},
  \bibinfo{journal}{JHEP} \textbf{\bibinfo{volume}{03}}, \bibinfo{pages}{118}
  (\bibinfo{year}{2018}), \eprint{1707.09298}.

\bibitem[{\citenamefont{Bousso and Polchinski}(2000)}]{Bousso:2000xa}
\bibinfo{author}{\bibfnamefont{R.}~\bibnamefont{Bousso}} \bibnamefont{and}
  \bibinfo{author}{\bibfnamefont{J.}~\bibnamefont{Polchinski}},
  \bibinfo{journal}{JHEP} \textbf{\bibinfo{volume}{06}}, \bibinfo{pages}{006}
  (\bibinfo{year}{2000}), \eprint{hep-th/0004134}.

\bibitem[{\citenamefont{Burgess}(2020)}]{Burgess:2020tbq}
\bibinfo{author}{\bibfnamefont{C.~P.} \bibnamefont{Burgess}},
  \emph{\bibinfo{title}{{Introduction to Effective Field Theory}}}
  (\bibinfo{publisher}{Cambridge University Press}, \bibinfo{year}{2020}), ISBN
  \bibinfo{isbn}{978-1-139-04804-0, 978-0-521-19547-8}.

\bibitem[{\citenamefont{Ferrara et~al.}(1994)\citenamefont{Ferrara, Kounnas,
  and Zwirner}}]{Ferrara:1994kg}
\bibinfo{author}{\bibfnamefont{S.}~\bibnamefont{Ferrara}},
  \bibinfo{author}{\bibfnamefont{C.}~\bibnamefont{Kounnas}}, \bibnamefont{and}
  \bibinfo{author}{\bibfnamefont{F.}~\bibnamefont{Zwirner}},
  \bibinfo{journal}{Nucl. Phys. B} \textbf{\bibinfo{volume}{429}},
  \bibinfo{pages}{589} (\bibinfo{year}{1994}), \bibinfo{note}{[Erratum:
  Nucl.Phys.B 433, 255--255 (1995)]}, \eprint{hep-th/9405188}.

\bibitem[{\citenamefont{Peskin and Schroeder}(1995)}]{Peskin:1995ev}
\bibinfo{author}{\bibfnamefont{M.~E.} \bibnamefont{Peskin}} \bibnamefont{and}
  \bibinfo{author}{\bibfnamefont{D.~V.} \bibnamefont{Schroeder}},
  \emph{\bibinfo{title}{{An Introduction to quantum field theory}}}
  (\bibinfo{publisher}{Addison-Wesley}, \bibinfo{address}{Reading, USA},
  \bibinfo{year}{1995}), ISBN \bibinfo{isbn}{978-0-201-50397-5}.

\end{thebibliography}

\end{document}